# Metadamping: An emergent phenomenon in dissipative metamaterials

Mahmoud I. Hussein and Michael J. Frazier

Department of Aerospace Engineering Sciences,

University of Colorado Boulder, Boulder, CO 80309

We theoretically demonstrate the concept of metadamping in dissipative metamaterials. We consider an infinite mass-spring chain with repeated local resonators and a statically equivalent periodic chain whose wave propagation characteristics are based on Bragg scattering. For each system we introduce identical viscous damping (dashpot) elements and compare the damping ratio associated with all Bloch modes. We find that the locally resonant metamaterial exhibits higher dissipation overall which indicates a damping emergence phenomena due to the presence of local resonance. We conclude our investigation by quantifying the degree of emergent damping as a function of the long-wave speed of sound in the medium or the static stiffness.

Damping is an intrinsic property of materials and its characterization is essential for quantifying the degree of energy dissipation or loss at different dynamic states. While in certain applications - such as acoustic sensing - dissipation is not desired, its maximization is sought in numerous avenues, including vibration suppression, shock resistance, and acoustic absorption.  A key trade-off in these latter applications, however, is that an increase in the intensity of damping commonly appears at the expense of stiffness, or mechanical load-bearing capacity.  For example, polymeric materials are generally more dissipative than metallic materials but are much less stiff.  Hence the discovery of routes leading to materials that are both highly dissipative and stiff is most advantageous.  Lakes et al. [1], for example, demonstrated that composite materials incorporating constituents in a metastable state simultaneously exhibit elevated values of viscoelastic stiffness and damping. Shape memory alloys also present another example where hysteretic motion of interfaces at the thermoelastic martensitic phase leads to realization of high damping capacity without reduction in overall stiffness [2]. Chung [3] presents a review of other approaches for enhancing damping-stiffness capacity of materials through composite materials engineering.

In this Letter, we focus on periodic acoustic metamaterials (AMs) with local resonance properties [4] as a candidate for materials that can be designed to exhibit high levels of dissipation while retaining high stiffness. The underlying principle is that vibration attenuation is mostly enhanced at resonance frequencies, as is well known in the field of structural dynamics. Here we draw an analogy and explore the effect of resonance on dissipation in the context of *material dynamics*. To assess the degree of dissipation in a locally resonant AM, we compare band structure characteristics with those of a corresponding statically-equivalent phononic crystal (PC), i.e., another type of periodic material that does not possess local resonances yet has the same long-wave propagation characteristics. In this analysis we seek to determine the possibility of an *emergence of damping* in the AM. The concept of emergence in complex systems theory and other disciplines stems from the wildly held notion that for certain systems "the whole is greater than the sum of its parts" [5]. Although emergence is usually associated with a lack of predictability, the concept may also be relevant to systems whose properties are predictable in principle, yet unforeseen *a priori* [6]. In relation to chemistry and materials science, often the emerging *whole* stems from the structure while the *parts* are associated with the composition. From this perspective, our aim is to theoretically investigate emergence in the context of how the internal structure, of a material, may lead to enhanced dissipation when compared to other materials with the same composition, or the same equivalent static properties.

By considering lumped masses, springs, and viscous damping (dashpot) elements, we construct a simple 1D model of a damped diatomic AM (represented by a "mass-in-mass" configuration [7] as shown in Fig. 1a), and for comparison we also examine a corresponding 1D model of a damped diatomic PC (represented by a "mass-and-mass" configuration as shown in Fig. 1b). This choice of simple lumped parameter models allows us to focus on the underlying damping emergence phenomenon free from the distraction of irrelevant system complexities. Considering unit cell periodicity for both cases, we apply the generalized form of Bloch's theorem [8-10],

$$u_\alpha^{j+n}(x, \kappa; t) = \widetilde{U}_\alpha e^{i(\kappa x + n\kappa a) + \lambda t},$$ (1)

which represents the displacement of mass $\alpha$ in the $(j + n)$th unit cell in the periodic chain, and where $\widetilde{U}_\alpha$ denotes wave amplitude, $a$ is the lattice constant, $\kappa$ is the wavenumber and $\lambda$ is a complex frequency function that permits wave attenuation in time. In the limiting case of no damping, $\lambda = \pm i\omega$, and the usual form of Bloch's theorem is recovered. Substituting Eq. (1) into the governing equations for both the AM and the PC yields the characteristic equation for each which is solved to obtain the frequency, $\omega_d(\kappa)$, and damping ratio, $\xi(\kappa)$, band structures from the roots $\lambda_s(\kappa) = -\xi_s(\kappa)\omega_s(\kappa) \pm$ $i\omega_d(\kappa)$, $s = 1,2$, where $s$ represents the branch number. The explicit form of the characteristic equation and the solution for each of the systems is provided in the Appendix. From the complex solution, $\lambda_s(\kappa)$, we directly extract $\omega_{d_s}(\kappa) = \mathrm{Im}[\lambda_s(\kappa)]$ and $\xi_s(\kappa) = -\mathrm{Re}[\lambda_s(\kappa)]/\mathrm{Abs}[\lambda_s(\kappa)]$ for each of the two dispersion branches.

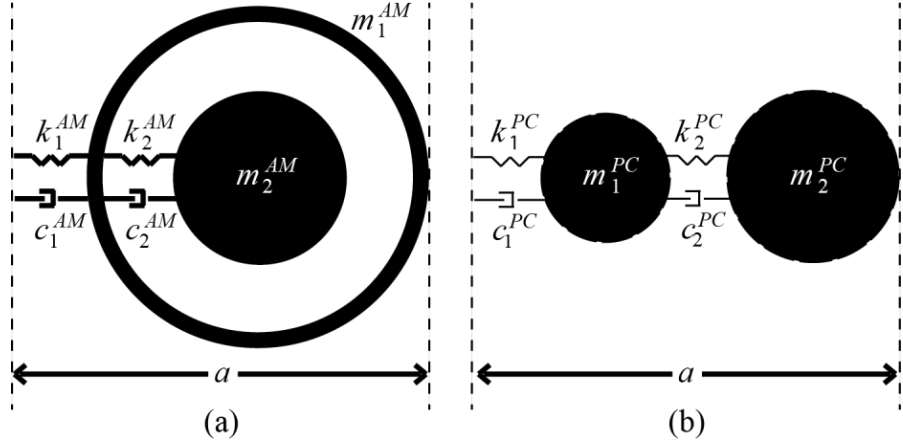

FIG 1. Unit cells of statically-equivalent periodic chains consisting of masses, springs and viscous damping (dashpot) elements: (a) acoustic metamaterial (mass-in-mass), (b) phononic crystal (mass-and-mass).

Now we examine the frequency and damping ratio dispersion curves of both systems for the following parameter ratios: $m_2^{PC}/m_1^{PC} = m_2^{AM}/m_1^{AM} = 5$, $k_2^{PC}/k_1^{PC} = k_2^{AM}/k_1^{AM} = 1/5$, and $c_2^{PC}/c_1^{PC} = c_2^{AM}/c_1^{AM} = 1$. Furthermore we set $m_1^{PC} = m_1^{AM} = 1$, $\omega_0^{PC} = \sqrt{k_2^{PC}/m_2^{PC}} = 100$, $a = 1$, and define $\eta = c_1^{PC} = c_2^{PC} = c_1^{AM} = c_2^{AM}$ (all the parameters here and elsewhere in this work may be read in SI units). The left hand side (LHS) of Fig. 2a displays the frequency band structure of the PC for three example cases: no damping ($\eta = 0$), low damping ($\eta = 40$), and high damping ($\eta = 80$), and the LHS of Figs. 2b and 2c, displays the damping ratio band structures for the low damping and high damping cases, respectively. On the right hand side (RHS) of Fig. 2, the matching diagrams for the corresponding AM are shown. In the damping ratio diagrams, we have added a third curve to represent the summation of the damping ratio values for the acoustic and optical branches, i.e., $\xi_{sum}^r(\kappa) = \xi_1^r(\kappa) + \xi_2^r(\kappa)$, where $r = PC$ or $AM$. To enable proper comparison, we select the value of $\omega_0^{AM} = \sqrt{k_2^{AM}/m_2^{AM}}$ in such a manner as to render both systems statically equivalent, that is, both systems having the same long-wave sound speed, or slope of the first frequency dispersion branch as

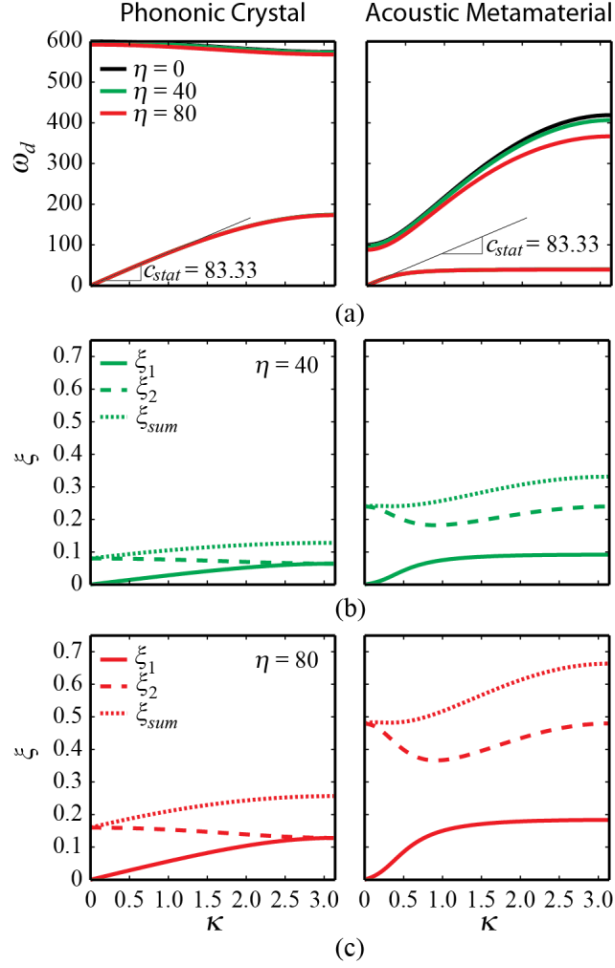

FIG 2 (color online). (a) Frequency band structure, and damping ratio band structure corresponding to (b) $\eta = 40$ and (c) $\eta = 80$, for statically equivalent PC (LHS) and AM (RHS).

the wavenumber tends to zero, $c_{stat} = \lim_{\kappa \to 0} \frac{d\omega_{d_1}}{d\kappa}$. With the chosen parameters for the PC, $c_{stat} = 83.33$. The AM is set to exhibit the same value of long-wave sound speed when $\omega_0^{AM} = 40.90$. The results show that there are shifts in the frequency band diagrams (particularly in the optical branches) due to the presence of damping, and that these shifts are more profound in the AM. This behavior manifests itself in a most remarkable manner in the damping ratio diagrams. Despite the static equivalence and equally prescribed value of the viscous damping constant, $\eta$, we observe in Figs. 2b and 2c that the AM exhibits higher damping ratio values (i.e., higher dissipation) across the entire Brillouin zone (BZ), for both the acoustic and optical branches. This is an indication of a considerable amplification, or emergence, of dissipation in the AM compared to its PC counterpart. To quantify this difference, we introduce a wavenumber-dependent *damping emergence metric*, $Z_l(\kappa) = \xi_l(\kappa)|_{AM} - \xi_l(\kappa)|_{PC}$, where $l = 1,2$ or $sum$. In Fig. 3, we show $Z_l(\kappa)$, its cumulative value,

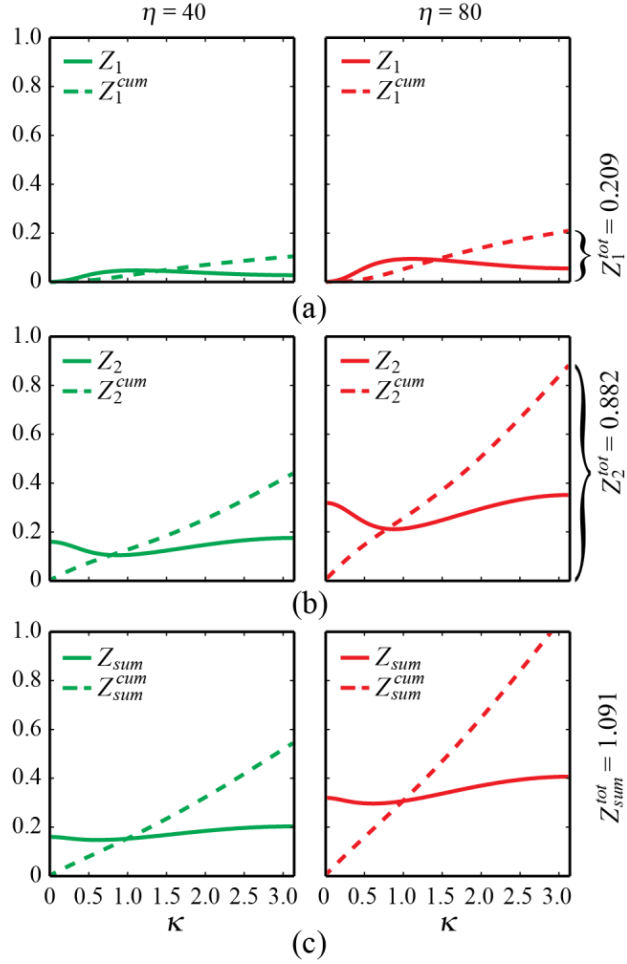

FIG. 3 (color online). Damping emergence metric, $Z_l$, for (a) acoustic branch ($l = 1$), (b) optical branch ($l = 2$), and (c) summation of the two branches ($l = sum$), for $\eta = 40$ (LHS) and $\eta = 80$ (RHS). The cumulative value, $Z_l^{cum}$, (as defined in Eq. 2) is also plotted, and the total value, $Z_{sum}^{tot}$, is marked for $\eta = 80$. All results are for $c_{stat} = 83.33$.

$$Z_l^{cum}(b) = \int_0^b Z_l \, d\kappa \, , \ l = 1,2 \text{ or } sum; \ b \in [0, \pi], \tag{2}$$

and its total value, $Z_l^{tot} = Z_l^{cum}(\pi)$ for our example PC and AM – displaying the data for the acoustic branch ($l = 1$), the optical branch ($l = 2$), and the summation of the two branches ($l = sum$) in Figs. 3a, 3b and 3c, respectively (for $\eta = 40$ on the LHS and $\eta = 80$ on the RHS). The results show significantly high values of $Z_l^{tot}$ which we may view as a measure of the intensity of damping emergence. We also observe that $Z_2 > Z_1$ for all values of $\kappa$. Fig. 4 shows that the value of $Z_l^{tot}$ varies linearly with the level of damping, $\eta$. Here the figure terminates when the value of the damping ratio of the AM optical branch is no longer limited to unity (i.e., critical damping) within the BZ. This occurs at $\eta = 165.57$. Returning to the driving objective of realizing materials that exhibit increased damping without deterioration of static stiffness, we repeat our calculations for a range of $c_{stat}$ values. We do so by keeping all the parameters as in the previous example except for $\omega_0^{PC}$. This causes $c_{stat}$ to vary, and keeping both systems statically equivalent, $\omega_0^{AM}$ varies accordingly. In Fig. 5, we show the relationship between the level of actual damping versus $c_{stat}$ for each of the PC and AM. For this purpose, we calculate the total damping ratio over both branches, $\xi_{sum}^{tot}$, which we obtain by integration, i.e., $\xi_{sum}^{tot} = \xi_{sum}^{cum}(\pi)$, where

$$\xi_{sum}^{cum}(b) = \int_0^b \xi_{sum}\,d\kappa, \text{ where } b \in [0, \pi]. \tag{3}$$

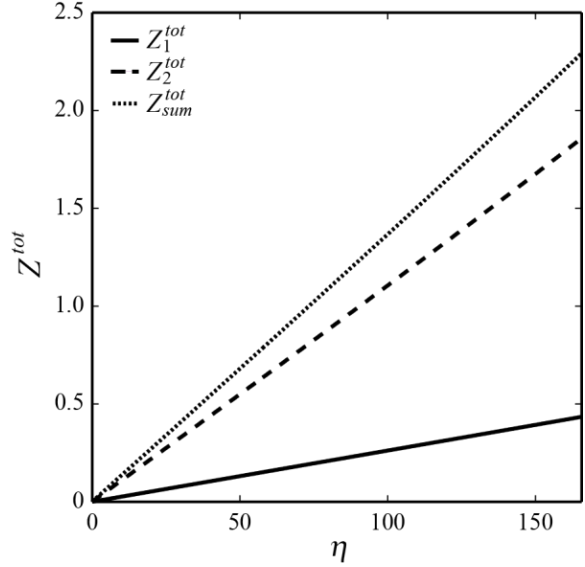

FIG 4. Total damping emergence, $Z_l^{tot}$, versus intensity of prescribed damping, $\eta$, for the acoustic branch ($l = 1$), optical branch ($l = 2$), and summation of the two branches ($l = sum$). Results are for $c_{stat} = 83.33$.

To provide further insight, we calculate an effective static Young's modulus, $E_{stat}$, which we obtain by considering an effective elastic rod, with a cross-sectional area equal to unity. Using the standard rod properties of $E$, the Young's modulus, $\rho$, the density, and $c = \sqrt{E/\rho}$, the speed of sound in the rod, we derive $E_{stat} \approx (m_1^r + m_2^r)c_{stat}^2$, where $r = PC$ or $AM$, for each of our periodic chains. We clearly observe in Fig. 5 that for a given value of long-wave speed, or effective static Young modulus, the AM exhibits a substantial increase in the value of $\xi_{sum}^{tot}$, when compared to the PC. Shaded in grey is the region of damping emergence, or *metadamping region*. We note that the level of metadamping (represented by the height of the shaded region) is highest at low levels of $c_{stat}$ (i.e., compliant

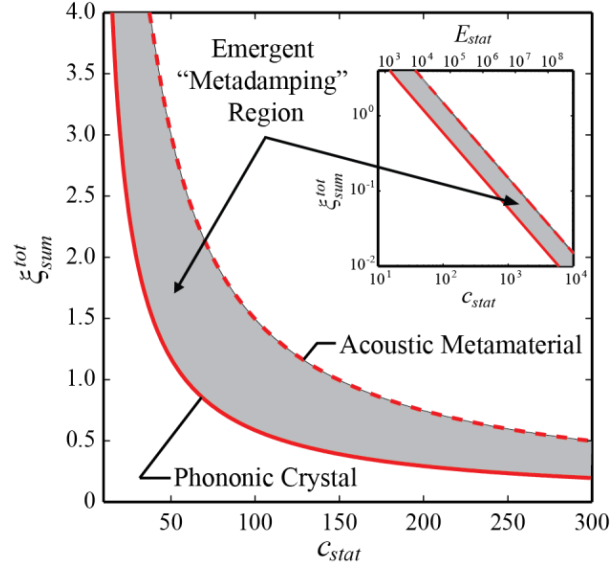

FIG 5 (color online). Graphical illustration of metadamping: total damping ratio over the two branches, $\xi_{sum}^{tot}$, versus long-wave speed of sound, $c_{stat}$, in the periodic chain for $\eta = 80$. The solid and dashed red lines correspond to PC and AM, respectively. The shaded grey region represents the region of damping emergence. In the inset the same plot is reproduced in log-log scale covering a wider range of $c_{stat}$ values. For further insight, another scale is added showing an effective static Young's modulus, $E_{stat}$, for an equivalent rod with unit cross-sectional area. The values of $E_{stat}$ are dependent on our selection of $m_1^{PC} = m_1^{AM} = 1$; upon selection of heavier masses, $E_{stat}$ will increase accordingly.

materials) and reduces in value as $c_{stat}$ increases to represent more stiff materials. The intensity of metadamping, or area of the metadamping region, increases with $\eta$ (as we may deduce from Fig. 4); however, for a given $\eta$, the region may be increased further upon optimization of the periodic chain parameters.

In conclusion, we have demonstrated the concept of damping emergence, or metadamping, due to the presence of local resonance. This finding has far reaching implications on the design of materials for numerous applications that require the reduction, mitigation, or absorption of vibrations, shock, or sound. While the analysis has been presented in the context of simple mass-spring-dashpot periodic chains, it can be readily extended to practical realizations of locally resonant acoustic metamaterials. Examples include configurations that utilize heavy inclusions with compliant coatings [4], soft inclusions [11], pillared structures [12,13], holey structures [14] and structures with suspended masses [15]. The two underlying features needed are (1) the presence of locally resonant elements and (2) the presence of at least one constituent material phase or component that exhibits damping (e.g., by utilizing viscoelastic materials, friction at material interphases etc.). It is the combination of these two features that may lead to metadamping. As such, other concepts for enhancing damping while retaining stiffness (such as those described in Ref. [1-3]) may be applied in conjunction with the inclusion of local resonators leading to an additive effect. Finally, while the concept of metadamping is presented here in the context of a mechanical problem, in principle it is applicable to other disciplines in materials physics that involve both resonance and dissipation.

## Acknowledgment

This work has been partially supported by the DARPA Young Faculty Award program (grant number: YFA N66001-11-1- 4134) and the NSF Graduate Research Fellowship program (award number: DGE 1144083).

## Appendix

### Dispersion relations for damped PC and AM mass-spring chains

Considering unit cell periodicity, the set of homogeneous equations describing the motion of each mass in the AM model shown in Fig. 1a is obtained as follows (where the AM superscript is omitted for brevity) [10]:

$$m_1\ddot{u}_1^j + c_1\left(2\dot{u}_1^j - \dot{u}_1^{j-1} - \dot{u}_1^{j+1}\right) + c_2\left(\dot{u}_1^j - \dot{u}_2^j\right) + k_1\left(2u_1^j - u_1^{j-1} - u_1^{j+1}\right) + k_2\left(u_1^j - u_2^j\right) = 0,$$
$$m_2\ddot{u}_2^j + c_2\left(\dot{u}_2^j - \dot{u}_1^j\right) + k_2\left(u_2^j - u_1^j\right) = 0,$$

(S1)

where $u_\alpha^j$ is the displacement of mass $\alpha$ in an arbitrary $j$th unit cell. In general, a unit cell and its neighbors may be identified by $j + n$, where $n = 0, -1, 1$ denote the present, previous, and subsequent unit cells, respectively. Similarly for the PC model shown in Fig. 1b, the equations of motion corresponding to the two masses are (where the PC superscript is omitted for brevity) [9]:

$$m_1\ddot{u}_1^j + (c_1 + c_2)\dot{u}_1^j - c_2\dot{u}_2^j - c_1\dot{u}_2^{j-1} + (k_1 + k_2)u_1^j - k_2u_2^j - k_1u_2^{j-1} = 0,$$
$$m_2\ddot{u}_2^j + (c_1 + c_2)\dot{u}_2^j - c_2\dot{u}_1^j - c_1\dot{u}_1^{j+1} + (k_1 + k_2)u_2^j - k_2u_1^j - k_1u_1^{j+1} = 0,$$

(S2)

For each of the system of equations, Eqs. (S1) and (S2), we apply the generalized form of Bloch's theorem [8-10] which is given in Eq. (1) and repeated here for convenience,

$$u_\alpha^{j+n}(x, \kappa; t) = \widetilde{U}_\alpha e^{i(\kappa x + n\kappa a) + \lambda t}.$$

Here, $u_\alpha^{j+n}$ represents the displacement of mass $\alpha$ in the $(j + n)$th unit cell in the periodic chain, and $\widetilde{U}_\alpha$ denotes wave amplitude, $a$ is the lattice constant, $\kappa$ is the wavenumber and $\lambda$ is a complex frequency function that permits wave attenuation in time. Upon substituting the Bloch solution into the equations of motion, we get a characteristic equation of the form

$$\lambda^4 + a\lambda^3 + b\lambda^2 + c\lambda + d = 0, \tag{S3}$$

where for the AM,

$$a = \frac{(m_1 + m_2)c_2 + 2m_2c_1(1 - \cos\kappa a)}{m_1 m_2},$$

$$b = \frac{(m_1 + m_2)k_2 + 2(m_2k_1 + c_1c_2)(1 - \cos\kappa a)}{m_1 m_2},$$

$$c = \frac{2(c_1k_2 + c_2k_1)(1 - \cos\kappa a)}{m_1 m_2}, \tag{S4}$$

$$d = \frac{2k_1k_2(1 - \cos\kappa a)}{m_1 m_2},$$

and for the PC,

$$a = \frac{(m_1 + m_2)(c_1 + c_2)}{m_1 m_2},$$

$$b = \frac{(m_1 + m_2)(k_1 + k_2) + 2c_1c_2(1 - \cos\kappa a)}{m_1 m_2},$$

$$c = \frac{2(c_1k_2 + c_2k_1)(1 - \cos\kappa a)}{m_1 m_2}, \tag{S5}$$

$$d = \frac{2k_1k_2(1 - \cos\kappa a)}{m_1 m_2}.$$

Upon solving Eq. (S3) for either the AM or PC, we obtain the general solution [9],

$$\lambda_{1,2} = \frac{1}{12}\left( 3a \mp \sqrt{9a^2 - 24b + 6\sqrt[3]{4}P + \frac{12\sqrt[3]{2}Q}{P}} \right.$$

$$\left. + \sqrt{18a^2 - 48b - 6\sqrt[3]{4}P - \frac{12\sqrt[3]{2}Q}{P} \mp \frac{54(a^3 - 4ab + 8c)}{\sqrt{9a^2 - 24b + 6\sqrt[3]{4}P + \frac{12\sqrt[3]{2}Q}{P}}}} \right) \tag{S6a}$$

where

$$Q = b^2 - 3ac + 12d, \tag{S6b}$$

$$P = \frac{}{\sqrt[3]{2b^3 - 9abc + 27(c^2 + a^2 d) - 72bd + \sqrt{-4Q^3 + [2b^3 - 9abc + 27(c^2 + a^2 d) - 72bd]^2}}}} \quad \text{(S6c)}$$

The distinction between the solution of the two systems arises upon the appropriate substitution of $a$, $b$, $c$, and $d$ as given in Eqs. (S4) and (S5), respectively.